\documentclass[11pt]{article}
\usepackage{amsmath}
\usepackage{graphicx}
\begin{document}
\title{Spin Fidelity for Three-qubit Greenberger-Horne-Zeilinger and W States Under Lorentz Transformations}
\author{Bahram Nasr Esfahani$^{1,}$\thanks{e-mail: ba$\_$nasre@sci.ui.ac.ir}
 , Mohsen Aghaee$^{2,}$\thanks{e-mail: mohsenaghaee1388@gmail.com}  }

\maketitle {\it \centerline{  $^1$Department of Physics, Faculty
of Sciences, University of Isfahan , Isfahan, Iran}} \maketitle
{\it \centerline{  $^2$Department of Physics, Faculty of Sciences,
Razi University , Kermanshah, Iran}}

\begin{abstract}

Constructing the reduced density matrix for a system of three
massive spin$-\frac{1}{2}$ particles described by a wave packet
with Gaussian momentum distribution and a spin part in the form of
GHZ or W state, the fidelity for the spin part of the system is
investigated from the viewpoint of moving observers in the jargon
of special relativity. Using a numerical approach, it turns out
that by increasing the boost speed, the spin fidelity decreases
and reaches to a non-zero asymptotic value that depends on the
momentum distribution and the amount of momentum entanglement.

\end{abstract}

\textbf{keywords}:Wigner rotation, spin density matrix, Gaussian
momentum distribution, fidelity,
GHZ state, W state.\\

\section{Introduction}

The role of special relativity in framing statements about quantum
information is illustrated by the fact that quantum entanglement
can depend on the reference frame of the observer. In practice,
Lorentz transformations can change the entanglement of the spins
of massive particles. Relativistic effects on quantum entanglement
and quantum information is investigated by many authors. One of
the early works in this area has considered a single free
spin-$\frac{1}{2}$ particle and by calculating the reduced density
matrix, it is shown that the spin entropy is not a relativistic
scalar \cite{peres}. Alsing and Milborn \cite{alsing} studied the
Lorentz transformation of maximally entangled Bell states. They
concluded that entanglement is Lorentz invariant.The entanglement
between the spins of a pair of particles may change because the
spin and momentum become mixed when viewed by a moving observer
\cite{ging}. Li an Du have investigated the quantum entanglement
between the spins of spin-$\frac{1}{2}$ massive particles in
moving frames, for the case that the momenta of the particles are
entangled \cite{li}. They have shown that, if the momenta of the
pair are appropriately entangled, the entanglement between the
spins of the Bell states remains maximal when viewed from any
Lorentz-transformed frame. Bartlett and Terno showed that
relativistically invariant quantum information can be encoded into
states of indistinguishable particles \cite{bartlett}. Recently,
simple examples have been presented of Lorentz transformation that
entangle the spins and momenta of two spin-$\frac{1}{2}$ particles
with positive mass such that no sum of entanglements have been
found to be unchanged \cite{jordan}. Fidelity for the spin part of
a system of two spin$-\frac{1}{2}$ particles described by a
Gaussian momentum distributed wave packet is studied from the view
point of moving observers and it is shown that the fidelity
decreases by increasing the boost velocity \cite{nasr}. Bell's
inequality in moving frames has been considered in several papers
\cite{czachor,ahn,lee,moon,kim,trashima}. The degree of violation
of Bell's inequality will decrease with increasing the velocity of
the observers if the directions of the measurements are fixed.
However, this doesn't imply a breakdown of nonlocal correlation
since the perfect anti-correlation is maintained in the
appropriately chosen different directions. Some efforts have been
done for extending these ideas to tripartite systems. For example,
Lorentz transformation of three-qubit GHZ state is studied and it
is shown that Bell's inequality is maximally violated for this
state \cite{moradi}. In tripartite discrete systems, two classes
of genuine tripartite entanglement have been discovered, namely,
the Greenberger-Horne-Zeilinger (GHZ) class \cite{ghz1,ghz2} and
the W class \cite{w,dur}. Some authors have provided proposals for
generation and observation of GHZ or W type entanglements
\cite{dik,jeong,rubin,sharma,song}.

In this paper we consider a moving system containing three
spin-$\frac{1}{2}$ massive particles such that in the rest frame,
its spin part be entangled as one of the GHZ or W states. In the
present approach, we introduce a Gaussian momentum distributed
wave packet represented in the momentum space for the system as
viewed in the rest frame. Also we introduce the wave packet for
the system as viewed by a boosted observer. Then we focus on the
spin part of each wave packet by finding the corresponding reduced
density matrix. As a result of relativistic spin decoherence, the
reduced density matrix observed in the boosted frame is mixed even
though it is prepared to be pure in the rest frame. We quantify
the amount of mixing via calculating the fidelity for these two
reduced density matrices. We will consider general boosts in the
$xz$-plane. To see the effect of momentum entanglement on the
results, the momentum part is chosen to be in extreme cases of
momentum product or momentum perfectly correlated.

\section{Lorentz transformation of reduced density matrix }

A three-particle quantum state is expressed by
\begin{equation}\label{bi}
       |\textbf{p}_1,\sigma_1;\textbf{p}_2,\sigma_2;\textbf{p}_3,\sigma_3\rangle=a^{\dagger}(\textbf{p}_1,\sigma_1)
       a^{\dagger}(\textbf{p}_2,\sigma_2)a^{\dagger}(\textbf{p}_3,\sigma_3)|\Phi_0\rangle,
\end{equation}
where $\textbf{p}$ is the 3-momentum vector, $\sigma$ is the spin
label, $a^{\dagger}$ is creation operator and $|\Phi_0\rangle$ is
the Lorentz invariant vacuum state. The state (\ref{bi}) has a
Lorentz transformation property \cite{weinberg} as
\begin{eqnarray}\label{bitr}
     U(\Lambda)|\textbf{p}_1,\sigma_1;\textbf{p}_2,\sigma_2;\textbf{p}_3,\sigma_3\rangle&=&\sum_{\sigma'_1\sigma'_2\sigma'_3}
      D_{\sigma'_1\sigma_1}(W(\Lambda,p_1))
      D_{\sigma'_2\sigma_2}(W(\Lambda,p_2))D_{\sigma'_3\sigma_3}(W(\Lambda,p_3))\\\nonumber
     &&\times|\Lambda\textbf{p}_{1},\sigma'_1;\Lambda\textbf{p}_{2},\sigma'_2;\Lambda\textbf{p}_{3},\sigma'_3\rangle,
\end{eqnarray}
where $\Lambda\textbf{p}$ is the spatial part of $\Lambda p$,
$D_{\sigma'\sigma}(W(\Lambda,p))$ is the unitary representation of
the Wigner rotation operator, and $W(\Lambda,p)$ is the Wigner's
little group element
\begin{equation}\label{a1}
      W(\Lambda,p)=L^{-1}(\Lambda p)\Lambda L(p),
\end{equation}
where $L(p)$ is the standard boost that takes a massive particle
of mass $m$ from rest to a 4-momentum $p$. The transformation of
the creation operator is as
\begin{equation}\label{a2}
      U(\Lambda)a^{\dagger}(\textbf{p},\sigma)U^{-1}(\Lambda)=\sum_{\sigma'}D_{\sigma'\sigma}(W(\Lambda,p))
      a^{\dagger}({\Lambda}\textbf{p},\sigma').
\end{equation}

Let $\hat{\textbf{p}}$ be the unit vector along the 3-momentum of
a particle as viewed in the rest frame, and consider a boost along
$\hat{\textbf{e}}$ with speed $V$. Then, the Wigner rotation
operator is represented as
\begin{equation}\label{a3}
       D(W(\Lambda,p))={\bf 1}\cos
       \frac{\Omega}{2}+i({\bf \sigma} \cdot\hat{\textbf{n}})\sin \frac{\Omega}{2},
\end{equation}
where ${\bf 1}$ is the unit $2\times 2$ matrix, ${\bf
\sigma}=\left(\sigma_1,\sigma_2,\sigma_3\right)$ denotes the Pauli
matrices,
$\hat{\textbf{n}}=\frac{\hat{\textbf{e}}\times\hat{\textbf{p}}}{|\hat{\textbf{e}}\times\hat{\textbf{p}}|}$
and
\begin{equation}\label{a4}
      \cot\frac{\Omega}{2}=\frac{\coth\frac{\xi}{2}\coth\frac{\eta}{2}
      +\hat{\textbf{e}}\cdot\hat{\textbf{p}}}{|\hat{\textbf{e}}\times\hat{\textbf{p}}|},
\end{equation}
where
\begin{equation}\label{a5}
      \cosh\xi=\frac{p^0}{m},\,\,\,\,\,\,\,\,\,\,\,\,\,\,\,\tanh\eta=\frac{V}{c}.
\end{equation}

In the rest frame of the observer, the 4-momentum of each particle
can be written in polar coordinates as
\begin{equation}
      p^{\mu}=[mc\,\cosh\xi, mc\sinh\,\xi(\cos\varphi\sin\vartheta,
      \sin\varphi\sin\vartheta, \cos\vartheta)],
\end{equation}
where $\tanh\xi=\frac{v}{c}$ and $v$ is the speed of the particle.
To be specific, we suppose that particles are moving along the
positive $x$-axis, i.e., $\vartheta=\frac{\pi}{2}$ and
$\varphi=0$, then the 4-momentum reduces to
\begin{equation}\label{a6}
      p^{\mu}=(mc\cosh\xi, mc\sinh\xi, 0, 0).
\end{equation}
This means that all of the three particles are assumed  move along
the positive $x$-axis, so for an arbitrary boost direction
$\hat{\textbf{e}}=(\cos\phi\sin\theta, \sin\phi\sin\theta,
\cos\theta)$, the axis of Wigner rotation $\hat{\textbf{n}}$ is
perpendicular to the $x$-axis in the same one direction for all
three particles. We will consider only $\phi=0$.

In this case the boost direction lies in the $xz$-plane and the
Wigner rotation is about the $y$-axis as
\begin{eqnarray}\label{a7}
     D(W(\Lambda,p))=\left(%
       \begin{array}{cc}
        D_{\uparrow\uparrow}(\Omega) & D_{\uparrow\downarrow}(\Omega) \\
        D_{\downarrow\uparrow}(\Omega) & D_{\downarrow\downarrow}(\Omega) \\
\end{array}%
   \right)=
   \left(%
   \begin{array}{cc}
   \cos\frac{\Omega}{2} & \sin\frac{\Omega}{2} \\
   -\sin\frac{\Omega}{2} & \cos\frac{\Omega}{2} \\
    \end{array}%
    \right)
\end{eqnarray}
where
\begin{equation}\label{a44}
      \cot\frac{\Omega}{2}=\frac{\coth\frac{\xi}{2}\coth\frac{\eta}{2}
      +\sin\theta}{\cos\theta},
\end{equation}

One may provide the argument using 3-particle states (\ref{bi})
and (\ref{bitr}) which require a sharp momentum distribution
around momentum $\textbf{p}$ for each particle. But the  realistic
situation involves a wave packet of the system with a definite
momentum distribution. We follow the argument in terms of Gaussian
momentum distributed wave packets.

In the rest frame of an observer the  wave packet in momentum
representation for the 3-particle system generally can be
expressed as
\begin{equation}\label{e1}
       |\psi\rangle=\sum_{\sigma_1 \sigma_2
          \sigma_3}\int\int\int d^3{\textbf{p}}_1 d^3{\textbf{p}}_2 d^3{\textbf{p}}_3
          \,g_{\sigma_1 \sigma_2
         \sigma_3}({\textbf{p}}_1,{\textbf{p}}_2,{\textbf{p}}_3)
       |{\textbf{p}}_1,\sigma_1;{\textbf{p}}_2,\sigma_2;{\textbf{p}}_3,\sigma_3\rangle,
\end{equation}
where
$g_{\sigma_1\sigma_2\sigma_3}({\textbf{p}}_1,{\textbf{p}}_2,{\textbf{p}}_3)$
is a distribution function for momentum and spin that is
normalized as
\begin{equation}\label{e2}
       \sum_{\sigma_1 \sigma_2 \sigma_3}\int\int\int
         d^3{\textbf{p}}_1 d^3{\textbf{p}}_2 d^3{\textbf{p}}_3 |g_{\sigma_1
         \sigma_2
       \sigma_3}({\textbf{p}}_1,{\textbf{p}}_2,{\textbf{p}}_3)|^2=1,
\end{equation}
which makes to have
\begin{equation}\label{e3}
       \langle{\textbf{p}}'_1,\sigma'_1;{\textbf{p}}'_2,\sigma'_2;{\textbf{p}}'_3,\sigma'_3|
        {\textbf{p}}_1,\sigma_1;{\textbf{p}}_2,\sigma_2;{\textbf{p}}_3,\sigma_3\rangle
         =\delta^3({\textbf{p}}'_1-{\textbf{p}}_1)
         \delta^3({\textbf{p}}'_2-{\textbf{p}}_2)\delta^3({\textbf{p}}'_3-{\textbf{p}}_3)\delta_{\sigma'_1
       \sigma_1}\delta_{\sigma'_2 \sigma_2}\delta_{\sigma'_3 \sigma_3}.
\end{equation}
Now, regarding (\ref{bitr}), for a boosted observer the  state
(\ref{e1}) changes to
\begin{eqnarray}\label{e4}
      |\psi^b\rangle&=&\sum_{\sigma_1 \sigma_2 \sigma_3}
         \sum_{\sigma'_1 \sigma'_2 \sigma'_3}\int d^3{\textbf{p}}_1 \int
          d^3{\textbf{p}}_2 \int d^3{\textbf{p}}_3 \sqrt{\frac{(\Lambda
           p_1)^0}{p_1^0}}\sqrt{\frac{(\Lambda
           p_2)^0}{p_2^0}}\sqrt{\frac{(\Lambda p_3)^0}{p_3^0}}g_{\sigma_1
           \sigma_2\sigma_3}({\textbf{p}}_1,{\textbf{p}}_2,{\textbf{p}}_3)\\\nonumber
           &&\times D_{\sigma'_1 \sigma_1}(W(\Lambda,p_1))D_{\sigma'_2
          \sigma_2}(W(\Lambda,p_2)) D_{\sigma'_3
         \sigma_3}(W(\Lambda,p_3))|\Lambda\textbf{p}_{1},\sigma'_1;\Lambda\textbf{p}_{2},\sigma'_2;
         \Lambda\textbf{p}_{3},\sigma'_3\rangle,
\end{eqnarray}
where $D_{\sigma'_{i}\sigma_i}$ is given by (\ref{a7}).

The density operators corresponding to these pure states are
$\rho=|\psi\rangle \langle \psi|$ and $ \rho^b=|\psi^b\rangle
\langle\psi^b|$, however we need the reduced density operators
$\varrho$ and $\varrho^b$ obtained by tracing over the momentum of
the density operators, that is
\begin{equation}\label{e5}
       \varrho_{\sigma'_1 \sigma'_2
        \sigma'_3,\sigma_1 \sigma_2 \sigma_3}=\int\int\int
         d^3{\textbf{p}}_1 d^3{\textbf{p}}_2 d^3{\textbf{p}}_3 \,g_{\sigma'_1
         \sigma'_2
        \sigma'_3}({\textbf{p}}_1,{\textbf{p}}_2,{\textbf{p}}_3)g^*_{\sigma_1
       \sigma_2 \sigma_3}({\textbf{p}}_1,{\textbf{p}}_2,{\textbf{p}}_3),
\end{equation}
and
\begin{eqnarray}\label{e6}
      \varrho^b_{\sigma'_1 \sigma'_2 \sigma'_3,\sigma_1
        \sigma_2 \sigma_3}&=&\sum_{\sigma''_1 \sigma''_2
          \sigma''_3}\sum_{\sigma'''_1 \sigma'''_2 \sigma'''_3}\int\int\int
           d^3{\textbf{p}}_1 d^3{\textbf{p}}_2
            d^3{\textbf{p}}_3\\\nonumber
             &&\times\left[D_{\sigma'_1 \sigma''_1}(W(\Lambda_1,p_1))
             D_{\sigma'_2\sigma''_2}(W(\Lambda_2,p_2)) D_{\sigma'_3
              \sigma''_3}(W(\Lambda_3,p_3))g_{\sigma''_1 \sigma''_2
              \sigma''_3}({\textbf{p}}_1,{\textbf{p}}_2,{\textbf{p}}_3)
              \right]\\\nonumber
            &&\times\left[D_{\sigma_1 \sigma'''_1}(W(\Lambda_1,p_1))
           D_{\sigma_2 \sigma'''_2}(W(\Lambda_2,p_2)) D_{\sigma_3
         \sigma'''_3}(W(\Lambda_3,p_3))g_{\sigma'''_1 \sigma'''_2
      \sigma'''_3}({\textbf{p}}_1,{\textbf{p}}_2,{\textbf{p}}_3)\right]^*.
\end{eqnarray}
Clearly $\varrho^b$ denotes the Lorentz transformed form of
$\varrho$. It must be noted that $\varrho^b$  will be mixed even
if $\varrho$ is pure.

Our purpose here is to calculate the fidelity of $\varrho$ and
$\varrho^b$, which accordingly we call it the spin fidelity $F_s$.
Then, we use the Uhlmann definition  \cite{uhlman,hubner} for
fidelity of mixed states, that is
\begin{equation}\label{e7}
       F_s=\left[\textrm{Tr}(\sqrt{\sqrt{\varrho}\, \varrho^b\sqrt{\varrho}})\right]^2.
\end{equation}
Fidelity is a basic ingredient in communication theory and for any
given communication scheme it is a quantitative measure of the
accuracy of the transmission. It takes numbers between 0 and 1; a
perfect communication corresponds to the fidelity 1. In the
present argument the spin fidelity quantify how $\varrho^b$ looks
like $\varrho$.

Recall that in our problem each 3-momentum has only one component
along the $x$-axis. Then all the three-dimensional integrals
reduce to one-dimensional integrals where the bold notation for
the momenta is suppressed. In the following $p$ stands for the
$x$-component of 3-momentum, not for the 4-momentum.

Now we choose the distribution function $g_{\sigma_1
\sigma_2\sigma_3}(p_1,p_2,p_3)$ such that the spin and the
momentum parts be separable and assume that the spin part is in
GHZ or W state. In tripartite discrete systems, two classes of
genuine tripartite entanglement have been discovered, namely, the
GHZ class \cite{ghz1,ghz2} and the W class \cite{w,dur}. These two
different types of entanglement are not equivalent and cannot be
converted to each other by local unitary operations combined with
classical communication. In terms of the spin basis, the GHZ state
has the form
$|\textrm{GHZ}\rangle=\frac{1}{\sqrt{2}}\left(|\uparrow\uparrow\uparrow\rangle+
|\downarrow\downarrow\downarrow\rangle\right)$ and the W state
takes the form
$|\textrm{W}\rangle=\frac{1}{\sqrt{3}}\left(|\uparrow\downarrow\downarrow\rangle+
|\downarrow\uparrow\downarrow\rangle+
|\downarrow\downarrow\uparrow\rangle\right)$. The entanglement in
the W state is robust against the loss of one qubit, while the GHZ
state is reduced to a product of two qubits. According to the
geometric measure of entanglement, the W state has higher
entanglement than the GHZ state does \cite{wei}. Methods are
proposed for generation and observation of GHZ or W type
entanglements \cite{sharma,song}. In our argument, it becomes
apparent that there is an interesting contrast between the
behavior of spin fidelity for these two states.

\section{GHZ state}

In this section we specify
$g_{\sigma_1\sigma_2\sigma_3}(p_1,p_2,p_3)$ such that in the rest
frame the spin part of the wave packet be in the GHZ state and, to
perceive the effect of momentum correlation on the results, the
momentum part be in states with different momentum correlation.

First we choose the distribution function as
\begin{equation}\label{e9}
      g_{\sigma_1 \sigma_2
       \sigma_3}(p_1,p_2,p_3)=f(p_1)
        f(p_2)
         f(p_3)\frac{1}{\sqrt{2}}\left(\delta_{\sigma_1 \uparrow}
        \delta_{\sigma_2 \uparrow} \delta_{\sigma_3
       \uparrow}+\delta_{\sigma_1 \downarrow}\delta_{\sigma_2
      \downarrow}\delta_{\sigma_3 \downarrow}\right),
\end{equation}
which as applied in (\ref{e1}), evidently gives a state that its
spin part is entangled as the GHZ state and the momentum part is
separable, i.e., the momentum entanglement is zero. We refer to
this choice as the momentum product case. For evaluating the
integrals, we should pick out a specified form for $f(p)$. Here we
consider it as
\begin{equation}\label{e10}
      f(p)=\frac{2}{(\alpha^2
       \pi)^\frac{1}{4}}\exp\left[-\frac{1}{2}\left(\frac{p}{\alpha}\right)^2\right],
\end{equation}
which shows a  Gaussian distribution (minimum uncertainty) of
momentum around $p=0$ with a width determined by $\alpha$.
Applying (\ref{e9}) in (\ref{e5}) and (\ref{e6}) we get
\begin{equation}\label{e11}
      \varrho=\frac{1}{2}
      \left(%
      \begin{array}{cccccccc}
      1 & 0 & 0 & 0 & 0 & 0 & 0 & 1 \\
      0 & 0 & 0 & 0 & 0 & 0 & 0 & 0 \\
      0 & 0 & 0 & 0 & 0 & 0 & 0 & 0 \\
      0 & 0 & 0 & 0 & 0 & 0 & 0 & 0 \\
      0 & 0 & 0 & 0 & 0 & 0 & 0 & 0 \\
      0 & 0 & 0 & 0 & 0 & 0 & 0 & 0 \\
      0 & 0 & 0 & 0 & 0 & 0 & 0 & 0 \\
      1 & 0 & 0 & 0 & 0 & 0 & 0 & 1 \\
      \end{array}%
      \right),
\end{equation}
which is pure and
\begin{eqnarray}\label{e12}
      \varrho^b_{\sigma'_1 \sigma'_2 \sigma'_3, \sigma_1 \sigma_2
       \sigma_3}&=&\frac{1}{2}\int d p_1 |f(p_1)|^2 \int d p_2
        |f(p_2)|^2 \int d p_3 |f(p_3)|^2\\\nonumber
         &&\times[D_{\sigma'_1\uparrow}(\Omega_1)D_{\sigma'_2 \uparrow}(\Omega_2)D_{\sigma'_3
          \uparrow}(\Omega_3)D^*_{\sigma_1 \uparrow}(\Omega_1)D^*_{\sigma_2
          \uparrow}(\Omega_2)D^*_{\sigma_3
           \uparrow}(\Omega_3)\\\nonumber
             &&+D_{\sigma'_1\uparrow}(\Omega_1)D_{\sigma'_2 \uparrow}(\Omega_2)D_{\sigma'_3
             \uparrow}(\Omega_3)D^*_{\sigma_1 \downarrow}(\Omega_1)D^*_{\sigma_2
             \downarrow}(\Omega_2)D^*_{\sigma_3
             \downarrow}(\Omega_3)\\\nonumber
             &&+D_{\sigma'_1
            \downarrow}(\Omega_1)D_{\sigma'_2
           \downarrow}(\Omega_2)D_{\sigma'_3 \downarrow}(\Omega_3)D^*_{\sigma_1
          \uparrow}(\Omega_1)D^*_{\sigma_2 \uparrow}(\Omega_2)D^*_{\sigma_3
         \uparrow}(\Omega_3)\\\nonumber
        &&+D_{\sigma'_1\downarrow}(\Omega_1)D_{\sigma'_2
       \downarrow}(\Omega_2)D_{\sigma'_3 \downarrow}(\Omega_3)D^*_{\sigma_1
      \downarrow}(\Omega_1)D^*_{\sigma_2 \downarrow}(\Omega_2)D^*_{\sigma_3
      \downarrow}(\Omega_3)],
\end{eqnarray}
which is mixed. Since $\varrho$ is pure $\varrho^2=\varrho$ and
(\ref{e7}) can be written as
\begin{equation}\label{ep7}
       F_s=\left[\textrm{Tr}(\sqrt{\varrho\, \varrho^b\varrho})\right]^2.
\end{equation}
Using (\ref{e11}) and (\ref{e12}) in  (\ref{ep7}), the spin
fidelity is obtained as
\begin{equation}\label{e13}
      F_s=\frac{1}{2}\left(\varrho^{b}_{\uparrow\uparrow\uparrow,\uparrow\uparrow\uparrow}+
       \varrho^{b}_{\uparrow\uparrow\uparrow,\downarrow\downarrow\downarrow}+
       \varrho^{b}_{\downarrow\downarrow\downarrow,\uparrow\uparrow\uparrow}
       +\varrho^{b}_{\downarrow\downarrow\downarrow,\downarrow\downarrow\downarrow}\right).
\end{equation}
We apply the Wigner rotation (\ref{a7}) in (\ref{e12}), then we
obtain
\begin{eqnarray}\label{e14}
       F_s&=&\frac{1}{8}[\,\overline{\cos\Omega_1}\,\,\overline{\cos\Omega_2}\,\,\overline{\cos\Omega_3}+
        \overline{\cos\Omega_1}\,\,\overline{\cos\Omega_2}+\overline{\cos\Omega_1}\,\,\overline{\cos\Omega_3}\\\nonumber
        &&+\overline{\cos\Omega_2}\,\,\overline{\cos\Omega_3}+\overline{\cos\Omega_1}+\overline{\cos\Omega_2}+
       \overline{\cos\Omega_3}+1],
\end{eqnarray}
where for each particle
\begin{equation}\label{e18}
        \overline{\cos\Omega}=\frac{2}{\gamma\sqrt{\pi}}\int_{0}^\infty d\xi
         \,e^{-\gamma^{-2}
         \sinh^2 \xi} \cosh\xi\left[1-\frac{\cos^2\theta\left(1+\cosh\eta\cosh\xi-\cosh\eta-\cosh\xi\right)}
        {1+\cosh\eta\cosh\xi+\sin\theta\sinh\eta\sinh\xi}\right],
\end{equation}
where $\gamma=\alpha/mc$, $\sinh\xi=p/mc$ and we have used
(\ref{a44}) and (\ref{e10}). Note that
$\overline{\cos\Omega_1}=\overline{\cos\Omega_2}=\overline{\cos\Omega_3}$,
then spin fidelity (\ref{e14}) reduces to
\begin{equation}\label{e16}
        F_s=\frac{1}{8}\left(\overline{\cos\Omega}^{\,\,3}+3\overline{\cos\Omega}^{\,\,2}+3
        \overline{\cos\Omega}+1\right).
\end{equation}

Next, we consider a case that two of three momenta, say $p_2$ and
$p_3$ are perfectly correlated. We refer to it as the 2-momentum
correlated case.  Thus we write
\begin{equation}\label{e20}
          g_{\sigma_1 \sigma_2 \sigma_3}(p_1,p_2,p_3)=
           f(p_1)f(p_2)\sqrt{\delta(p_2-p_3)}
            \frac{1}{\sqrt{2}}\left(\delta_{\sigma_1 \uparrow} \delta_{\sigma_2
           \uparrow} \delta_{\sigma_3 \uparrow}+\delta_{\sigma_1 \downarrow}
          \delta_{\sigma_2 \downarrow} \delta_{\sigma_3 \downarrow}\right),
\end{equation}
which as substituted in (\ref{e5}) and (\ref{e6}) gives a pure
$\varrho$ as (\ref{e11}) and a mixed $\varrho^b$. Applying the
results in (\ref{ep7}), we have
\begin{equation}\label{e21}
          F_s=\frac{1}{8}\left(\,\,\overline{\cos\Omega_1}\,\,\overline{\cos^2\Omega_2}+
           2\,\,\overline{\cos\Omega_1}\,\,\overline{\cos\Omega_2}
           +\overline{\cos^2\Omega_2}+\overline{\cos\Omega_1}+2\,\,\overline{\cos\Omega_2}+
          1\right),
\end{equation}
which is comparable with (\ref{e14}). Again note that
$\overline{\cos\Omega_1}=\overline{\cos\Omega_2}$, then
\begin{equation}\label{e22}
        F_s=\frac{1}{8}\left(\overline{\cos\Omega}\,\,\overline{\cos^2\Omega}+2\,\,\overline{\cos\Omega}^{\,\,2}+
        \overline{\cos^2\Omega}+3\overline{\cos\Omega}+1\right),
\end{equation}
where
\begin{equation}\label{e08}
        \overline{\cos^2\Omega}=\frac{2}{\gamma\sqrt{\pi}}\int_{0}^\infty d\xi
         \,e^{-\gamma^{-2}
         \sinh^2 \xi} \cosh\xi\left[1-\frac{\cos^2\theta\left(1+\cosh\eta\cosh\xi-\cosh\eta-\cosh\xi\right)}
        {1+\cosh\eta\cosh\xi+\sin\theta\sinh\eta\sinh\xi}\right]^2.
\end{equation}

Here we note that, in (\ref{e20}) as well as in the following
arguments, the delta functions should be regarded as limits of
analytical functions under certain conditions. Precisely,
perfectly correlated momenta should be regarded as a limiting case
of entangled Gaussian momenta \cite{li}. An experimental situation
for generating the momentum entanglement is discussed by Lamata
{\it et.al.} \cite{solano}. They studied the dynamics of momentum
entanglement generated in the lowest-order QED interaction between
two massive spin-$\frac{1}{2}$ charged particles, which grows in
time as the two fermions exchange virtual photons. In this scheme
the degree of generated entanglement between interacting particles
with initial well-defined momentum can be infinite, however, they
explained this divergence in the context of entanglement theory
for continuous variables, and showed how to circumvent this
apparent paradox.

In order to have all the three momenta perfectly correlated (
3-momentum correlated case) we choose
\begin{eqnarray}\label{e24}
        g_{\sigma_1 \sigma_2
         \sigma_3}(p_1,p_2,p_3)&=&
          f(p_1)\sqrt{\delta(p_1-p_2)\delta
          (p_1-p_3)}\\\nonumber
          && \times \frac{1}{\sqrt{2}}(\delta_{\sigma_1 \uparrow} \delta_{\sigma_2
         \uparrow} \delta_{\sigma_3 \uparrow}+\delta_{\sigma_1 \downarrow}
        \delta_{\sigma_2 \downarrow} \delta_{\sigma_3 \downarrow}),
\end{eqnarray}
which as substituted in (\ref{e5}) and (\ref{e6}), gives
\begin{equation}\label{e26}
       F_s=\frac{1}{8}\left(\overline{\cos^3\Omega}+3\overline{\cos^2\Omega}+3
        \overline{\cos\Omega}+1\right),
\end{equation}
where
\begin{equation}\label{e008}
        \overline{\cos^3\Omega}=\frac{2}{\gamma\sqrt{\pi}}\int_{0}^\infty d\xi
         \,e^{-\gamma^{-2}
         \sinh^2 \xi} \cosh\xi\left[1-\frac{\cos^2\theta\left(1+\cosh\eta\cosh\xi-\cosh\eta-\cosh\xi\right)}
        {1+\cosh\eta\cosh\xi+\sin\theta\sinh\eta\sinh\xi}\right]^3,
\end{equation}

There is no analytical solution for the integrals (\ref{e18}),
(\ref{e08}) and (\ref{e008}) hence we switch to a numerical
approach for evaluating the spin fidelity $F_s$. It reveals that
the behavior of $F_s$ in terms of the boost velocity is in the
same form for all the $\theta$'s and the most change occurs for
$\theta=0$, that is, when the boost is along the $z$-axis.
Therefore, we proceed with substituting $\theta=0$ in the
integrals. Fig. 1 shows $F_s$ plotted numerically in terms of the
boost parameter $\eta$ for a given width for the momentum
distribution. The solid curve shows $F_s$ given by (\ref{e16}) in
the momentum product case. The dashed curve is plotted for $F_s$
given by (\ref{e21}) in the 2-momentum correlated case and the
dashed-dotted curve is for $F_s$ given by (\ref{e26}) in the
3-momentum correlated case. We see that by increasing the boost
velocity more spin decoherence occurs and expectedly the spin
fidelity decreases with increasing $\eta$. By increasing the
momentum correlation, $F_s$ decreases less, such that for small
$\eta$ the curves coincide, and for $\eta\rightarrow\infty$ (ultra
relativistic limit) they slightly split to non-zero asymptotic
values. It can be shown that by decreasing the width $\gamma$, the
spin fidelity becomes less sensitive to $\eta$, hence the slope of
the curves decreases.

\begin{figure}
   \begin{center}
         \includegraphics[width=10cm,height=10cm,angle=0]{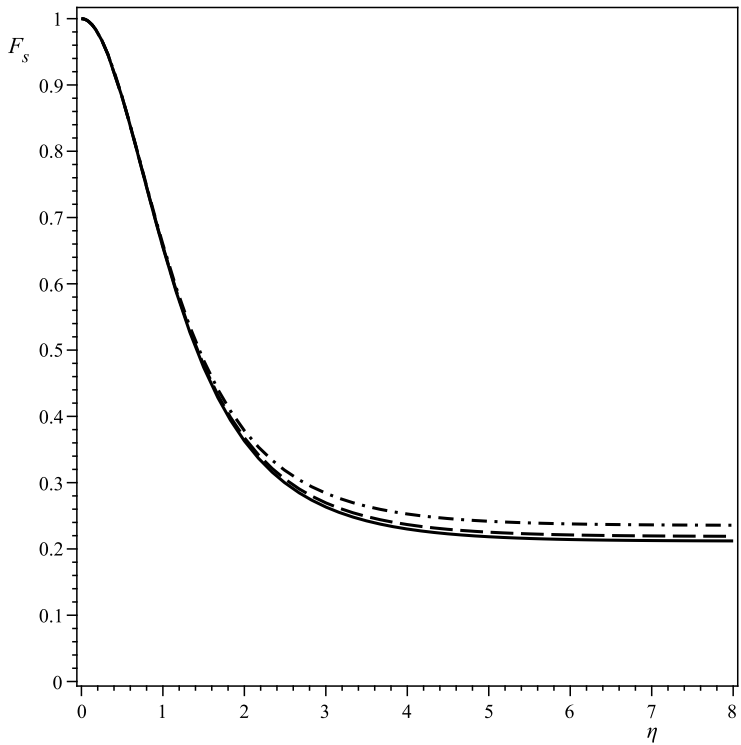}
         \caption{Spin fidelity $F_s$ versus the boost parameter $\eta$ in the GHZ
         case plotted for $\gamma=20$. The solid curve
         is plotted for the momentum product (zero momentum entanglement)
         while the dashed and the dashed-dotted correspond to the
         2-momentum correlated  and 3-momentum correlated cases,
         respectively. For small $\eta$ the curves coincide
         however at large $\eta$ they slightly split. }
   \end{center}
\end{figure}

\section{W state}

As the previous section we investigate the W state in three cases
of different momentum correlation. We begin with the momentum
product case, by choosing
\begin{equation}\label{e28}
       g_{\sigma_1 \sigma_2
        \sigma_3}(p_1,p_2,p_3)=f(p_1)
         f(p_2)f(p_3)\frac{1}{\sqrt{3}} \left(\delta_{\sigma_1
         \uparrow} \delta_{\sigma_2 \downarrow} \delta_{\sigma_3
          \downarrow}+\delta_{\sigma_1 \downarrow}\delta_{\sigma_2
         \uparrow}\delta_{\sigma_3 \downarrow}+\delta_{\sigma_1
        \downarrow}\delta_{\sigma_2 \downarrow}\delta_{\sigma_3
       \uparrow} \right).
\end{equation}
By this choice (\ref{e5}) leads to
\begin{equation}\label{e29}
       \varrho=\frac{1}{3}
       \left(%
       \begin{array}{cccccccc}
       0 & 0 & 0 & 0 & 0 & 0 & 0 & 0 \\
       0 & 0 & 0 & 0 & 0 & 0 & 0 & 0 \\
       0 & 0 & 0 & 0 & 0 & 0 & 0 & 0 \\
       0 & 0 & 0 & 1 & 0 & 1 & 1 & 0 \\
       0 & 0 & 0 & 0 & 0 & 0 & 0 & 0 \\
       0 & 0 & 0 & 1 & 0 & 1 & 1 & 0 \\
       0 & 0 & 0 & 1 & 0 & 1 & 1 & 0 \\
       0 & 0 & 0 & 0 & 0 & 0 & 0 & 0 \\
       \end{array}%
       \right),
\end{equation}
which is pure and  (\ref{e6}) gives
\begin{eqnarray}\label{e30}
      && \varrho^b_{\sigma'_1 \sigma'_2 \sigma'_3,\sigma_1 \sigma_2 \sigma_3}=\frac{1}{3}\int d p_1 |f(p_1)|^2
        \int d p_2 |f(p_2)|^2 \int d p_3 |f(p_3)|^2 \\\nonumber
         &&\times[D_{\sigma'_1\uparrow}(\Omega_1)D_{\sigma'_2 \downarrow}(\Omega_2)D_{\sigma'_3
          \downarrow}(\Omega_3)+D_{\sigma'_1\downarrow}(\Omega_1)D_{\sigma'_2 \uparrow}(\Omega_2)D_{\sigma'_3
           \downarrow}(\Omega_3)+D_{\sigma'_1\downarrow}(\Omega_1)D_{\sigma'_2
            \downarrow}(\Omega_2)D_{\sigma'_3\uparrow}(\Omega_3)]\\\nonumber
            &&\times[D_{\sigma_1\uparrow}(\Omega_1)D_{\sigma_2 \downarrow}(\Omega_2)D_{\sigma_3
           \downarrow}(\Omega_3)+D_{\sigma_1 \downarrow}(\Omega_1)D_{\sigma_2
         \uparrow}(\Omega_2)D_{\sigma_3 \downarrow}(\Omega_3)+D_{\sigma_1
        \downarrow}(\Omega_1)D_{\sigma_2 \downarrow}(\Omega_2)D_{\sigma_3
       \uparrow}(\Omega_3)]^*\nonumber.
\end{eqnarray}
Again apply these operators in (\ref{ep7}) and after doing some
manipulation the spin fidelity is calculated as
\begin{equation}\label{e31}
         F_s=\frac{1}{3}\left(\varrho^{b}_{\uparrow\downarrow\downarrow,\uparrow\downarrow\downarrow}+
          \varrho^{b}_{\uparrow\downarrow\downarrow,\downarrow\uparrow\downarrow}+
          \varrho^{b}_{\downarrow\uparrow\downarrow,\uparrow\downarrow\downarrow}+
           \varrho^{b}_{\uparrow\downarrow\downarrow,\downarrow\downarrow\uparrow}+
           \varrho^{b}_{\downarrow\downarrow\uparrow,\uparrow\downarrow\downarrow}+
            \varrho^{b}_{\downarrow\uparrow\downarrow,\downarrow\uparrow\downarrow}+
          \varrho^{b}_{\downarrow\uparrow\downarrow,\downarrow\downarrow\uparrow}+
          \varrho^{b}_{\downarrow\downarrow\uparrow,\downarrow\uparrow\downarrow}+
        \varrho^{b}_{\downarrow\downarrow\uparrow,\downarrow\downarrow\uparrow}\right),
\end{equation}
which as evaluated by (\ref{a7}), becomes
\begin{eqnarray}\label{e32}
       F_s&=&\frac{1}{72}\,\,[21\,\,\overline{\cos\Omega_1}\,\,\overline{\cos\Omega_2}\,\,\overline{\cos\Omega_3}+
         5\,\,\overline{\cos\Omega_1}\,\,\overline{\cos\Omega_2}+5\,\,\overline{\cos\Omega_1}\,\,\overline{\cos\Omega_3}\\\nonumber
         &&+5\,\,\overline{\cos\Omega_2}\,\,\overline{\cos\Omega_3}+5\,\,\overline{\cos\Omega_1}+5\,\,\overline{\cos\Omega_2}+
       5\,\,\overline{\cos\Omega_3}+21],
\end{eqnarray}
which reduces to
\begin{equation}\label{e33}
       F_s=\frac{1}{24}\,\,\left(7\,\,\overline{\cos\Omega}^{\,\,3}+5\,\,\overline{\cos\Omega}^{\,\,2}+5\,\,
       \overline{\cos\Omega}+7\right).
\end{equation}

Then, let
\begin{equation}\label{e34}
      g_{\sigma_1 \sigma_2\sigma_3}(p_1,p_2,p_3)=
        f(p_1)f(p_2)\sqrt{\delta(p_2-p_3)}
         \frac{1}{\sqrt{3}}\left(\delta_{\sigma_1 \uparrow} \delta_{\sigma_2
          \downarrow} \delta_{\sigma_3 \downarrow}+\delta_{\sigma_1
         \downarrow} \delta_{\sigma_2 \uparrow} \delta_{\sigma_3
        \downarrow}+\delta_{\sigma_1 \downarrow} \delta_{\sigma_2
      \downarrow} \delta_{\sigma_3 \uparrow}\right),
\end{equation}
which apparently describes the 2-momentum correlated case. After
doing some manipulations this leads to the following expression
for the spin fidelity
\begin{equation}\label{e36}
        F_s=\frac{1}{72}\left(39\,\,\overline{\cos\Omega}\,\,\overline{\cos^2\Omega}+
          7\,\,\overline{\cos^2\Omega}+10\,\,\overline{\cos\Omega}^{\,\,2}-3\,\,\overline{\cos\Omega}
        +19\right).
\end{equation}

Finally, we consider the 3-momentum correlated case designated by
\begin{eqnarray}\label{e37}
        g_{\sigma_1
          \sigma_2\sigma_3}(p_1,p_2,p_3)&=&
           f(p_1)\sqrt{\delta(p_1-p_2)\delta(p_1-p_3)}\\\nonumber
            &&\times\frac{1}{\sqrt{3}}(\delta_{\sigma_1 \uparrow} \delta_{\sigma_2
            \downarrow} \delta_{\sigma_3 \downarrow}+\delta_{\sigma_1
           \downarrow} \delta_{\sigma_2 \uparrow} \delta_{\sigma_3
          \downarrow}+\delta_{\sigma_1 \downarrow} \delta_{\sigma_2
         \downarrow} \delta_{\sigma_3 \uparrow} ).
\end{eqnarray}
Using this, we find the spin fidelity as
\begin{equation}
        F_s=\frac{1}{72}\left(75\,\,\overline{\cos^3\Omega}+25
        \,\,\overline{\cos^2\Omega}-39\,\,\overline{\cos\Omega}+11\right).
\end{equation}

Again the most change in these fidelities occurs for $\theta=0$ in
the integrals and we present the result for this case. The curves
in Fig. 2 are plotted numerically and describe the behavior of
$F_s$ in the present case in terms of the boost parameter $\eta$
for a given width $\gamma$. As is indicated, the solid curve, the
dashed curve and the dashed-dotted curve are plotted for the
momentum product case, the 2-momentum correlated case and the
3-momentum correlated case, respectively. Comparing with Fig. 1,
we conclude that the $F_s$ again descends to nonzero asymptotic
values but there is a significant separation between the three
curves. This means that the present W case is more sensitive to
the momentum entanglement. Also, note that the order of curves in
Fig. 2 is inverse of the order of curves in Fig. 1.

\begin{figure}
   \begin{center}
         \includegraphics[width=10cm,height=10cm,angle=0]{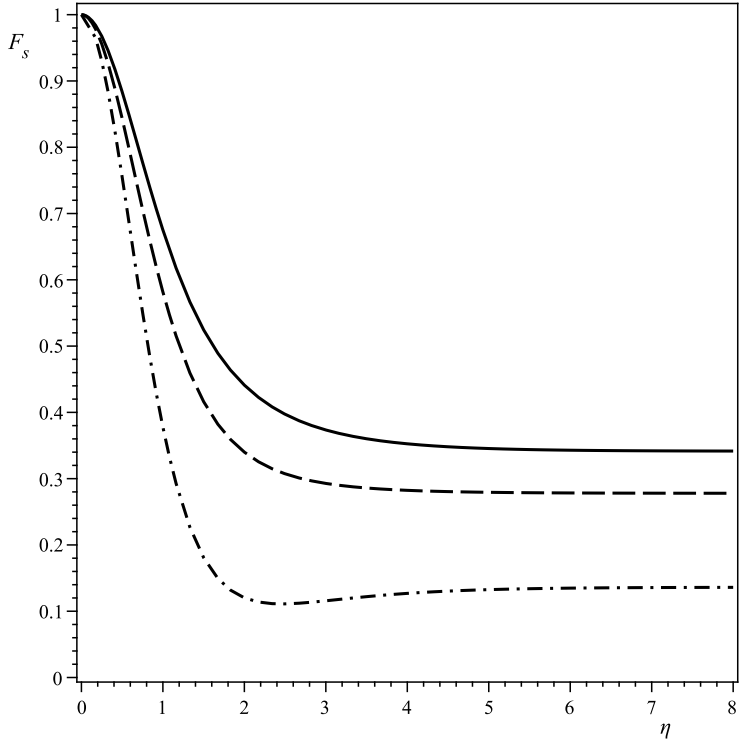}
         \caption{$F_s$ versus $\eta$ in the W
         case plotted for $\gamma=20$, for the boosts along the $z$-axis. The solid curve
         is plotted for the momentum product (zero momentum entanglement)
         while the dashed curve and the dashed-dotted curve correspond to the
         2-momentum correlated  and 3-momentum correlated cases,
         respectively.}
   \end{center}
\end{figure}

\section{Conclusions}

In this work we investigated  a system of three massive particles
described by a Gaussian momentum distributed wave packet such that
in the rest frame its spin part was entangled as the GHZ state or
the W state. Then we constructed the wave packet of the system as
viewed by a boosted observer, by using the corresponding Wigner
rotation operators. We focused on the spin part of the system by
tracing out the momentum part and  finding the reduced density
operators both for the rest observer and the boosted observer.
Using these reduced density matrices and the Uhlmann formula for
fidelity, the spin fidelities were formulated separately when
there was no momentum correlation, when two of three momenta were
correlated and when all the three momenta were correlated. We
could not evaluate fidelities analytically, so we utilized a
numerical approach to plot $F_s$ in terms of the boost parameter
$\eta$, as Fig.s 1 and 2 show.

We conclude that for the GHZ case, by increasing the boost
velocity, $F_s$ falls to non-zero asymptotic values that increase
as the momenta become more entangled. One may explain this
behavior by regarding the results of the refs. \cite{ging} and
\cite{li}. By boosting the wave packet, we move some of the spin
entanglement to the momentum part and simultaneously the momentum
entanglement appears to to be moved to the spins. The amount of
transferred entanglement grows with increasing the boost velocity.
Tracing out the momentum from the Lorentz-transformed density
matrix destroys some of the entanglement. This process causes to
decrease the spin entanglement in the boosted frame. When the
momenta are correlated, the transfer of momentum entanglement to
spins compensates somewhat the decrease of spin entanglement and
then the spin fidelity decreases less. However, as the figures
show, this becomes more significant at large boost velocities. For
the W case, the situation is inverse and by increasing the
momentum entanglement, the spin fidelity decreases more.

\end{document}